# 4. Scientometrics and the evaluation of European integration

**Koen Frenken and Loet Leydesdorff**

## 1. INTRODUCTION

A fascinating facet of the European Union is the multiplicity of meanings that are generated with respect to European integration. Different perceptions among Europeans contest the meaning of integration in general and regarding each social subsystem (economy, politics, science, sports, etc.) in particular. It is therefore both surprising and understandable that relatively few attempts have been reported to measure European integration in a formal way. This is surprising because of the importance of the question, but understandable because of the changing and conflicting meanings of European integration (Leydesdorff, 1992; Luukkonen, 1998).

In this chapter, we elaborate on the topic of European integration in science. We will not deal with questions related to the effects of European integration, but only with the scientometric question how one can quantitatively indicate integration of the European science system. This research question is in a certain sense a *sine qua non* for further research. Without indicators of integration, both the determinants and the effects of (European) integration are hard to assess statistically, let alone the question of the effectiveness of European science policies. Admittedly, however, the empiricist approach looses an explicit perspective of multiplicity of local meanings of "Europe". In this respect, our study is intended to facilitate and supplement debates rather than to provide a final answer to the questions whether European integration "exists".

In this chapter, we first discuss the use of scientometric indicators in research evaluation from a historical perspective in (section 2). A discussion of European science policy follows (section 3). Then, we introduce a number of indicators of integration and discuss our empirical results concerning the evolution of the European science system in the 1980s and 1990s (section 4). We close the chapter with a discussion of possible avenues of future research for enhancing research evaluation (section 5).

## 2. SCIENTOMETRICS AND RESEARCH EVALUATION

### The Endless Frontier

The idea that scientific knowledge could be organised deliberately and controlled from a mission perspective can be considered as resulting from experiences in World War II. Before that time the intellectual organisation of knowledge had largely been left to the internal mechanisms of discipline formation and specialist communications (Bush, 1945; Whitley, 1984). The military impact of science and technology through knowledge-based development and mission-oriented research during World War II (e.g., the Manhattan project) made it necessary in 1945 to formulate a new science and technology policy under peacetime conditions.

Vannevar Bush's report to the U.S. President entitled *The Endless Frontier* contained a plea for a less interventionist science policy (Bush, 1945). Quality control should be left to the internal mechanisms of the scientific elite, for example, through the peer review system. The model of the U.S. National Science Foundation (1947)[1] was thereafter followed by other Western countries. For example, the Netherlands created its foundation for Fundamental Scientific Research (ZWO) in 1950. With hindsight, one can consider this period as the institutional phase of science policies: the main policy instrument was the support of science with institutions to control its funding.

Alongside the military coordination by NATO, the *Organization for Economic Co-operation and Development* (OECD) was created in 1960 in order to organise science and technology policies among its member states. This led in 1963 to the *Frascati Manual for the Measurement of Scientific and Technical Activities* (1963), which can be understood as a response to the increased economic importance of science and technology. This manual defined parameters for the statistical monitoring of science and technology on a comparative basis. One was then able to compare the output performance and resource efficiency of various nation states. This rapidly led to questions concerning "strengths and weakness" of nations in specific disciplines and later led to policies based on differential increases in the budgets of particular disciplines. Thus, the focus remained on financial input-indicators, while the system relied on peer review in scientific disciplines for more detailed decision-making at the lower levels of individual scientific disciplines and specialties (Leydesdorff, 2003).

[1] Actually, President Truman vetoed the first N.S.F. act of 1947. The creation was then postponed until 1950.

## Output indicators

The use of scientometric indicators in research evaluation emerged in the 1970s in the United States and somewhat later in European countries. Before that time research evaluation proceeded mainly through the peer review system, on the one hand, and through economic indicators, on the other. The latter types of indicators (*e.g.*, percentage of GNP spent on R&D) have been developed by the *Organization of Economic Co-operation and Development* (OECD) in Paris, and can be considered as input indicators.

The *Science Citation Index* produced by Eugene Garfield's Institute of Scientific Information came to be recognised as a means to objectify standards using literature-based indicators (Price, 1963). The gradual introduction of output indicators such as number of publications and citations has proven socially legitimated both internally and externally to the science system. Internal use primarily consists of quality control and bench marking within and across disciplinary frameworks. For example, output records are increasingly used as a tool in the academic labour market. Externally, output indicators are used mainly by policy makers and science administrators who wish to assess institutions and to evaluate investments in research projects. In the early 1980s, scientometricians developed a fine-grained model that introduced output as a feedback parameter into the finance scheme of departments during the early 1980s (Moed *et al.*, 1985).

In different European countries, very different trajectories in the use of output indicators emerged. In some countries like the U.K., the idea of using output indicators to feedback on budgets was rapidly introduced as a tool in the funding of university research. The other European countries did not follow the UK in this extreme rationalisation of a budget model for research, but pressures prevailed during the 1990s to make publication and citation rates visible in evaluation exercises. For example, after the German unification in 1990, extensive evaluation of the research portfolio of Eastern Germany was immediately placed on the relevant scientific and policy agendas (Weingart, 1991).

**Methodological complications**

Publication and citation analyses have become standard tools for research evaluation. However, some methodological problems remain unresolved. The consequent uncertainties have been reflected in hesitations to apply these tools as standards in policy making processes and research management decisions. How shaky is the ground on which the evaluations stand?

First, one can legitimately raise the question of the unit of analysis in scientific knowledge production and control (Collins, 1985). The intellectual organisation of the sciences does not coincide with their institutional organisation. Scientist self-organise in communities that cross the institutional and national boundaries, while the budgetary organisation has remained largely within departments and nations states.

Second, the complex relationship between the intellectual and institutional organisation of research is especially problematic when dealing with emerging fields of research. New scientific developments (e.g., artificial intelligence) start in very different and unstable institutional settings. This calls for a *cognitive* unit of analysis rather than an institutional unit of analysis. However, cognitions cannot easily be observed or measured though progress in scientometrics is being made. One way to define a cognitive unit of analysis is to cluster journal-journal citations as citing relations reflect cognitive linkages (Leydesdorff and Cozzens, 1993; Van den Besselaar and Leydesdorff, 1996). We return to this issue in the final section of this chapter.

## 3. EUROPEANISATION AND S&T POLICY

**Subsidiarity**

The *Single Act* of the European Community in 1986 and the *Maastricht Treaty* of the European Union in 1991 have marked a gradual transition within Europe to a supra-national science, technology, and innovation policy. The EU policies continuously referred to science and technology (S&T), because these are considered as the strongholds of the common heritage of the member states. However, the 'subsidiarity' principle prescribes that the European Commission should not intervene in matters that can be left to the nation states. Therefore, a 'federal' research program of the EU could not be developed without taking the detour of a focus on *science-based innovation* using framework programmes rather than on basic science within a ongoing open-call programme (Narin and Elliott, 1985).

The national orientation on basic science and the European orientation of science-based innovation, explains why by far the largest share of European funding of science is still organised at the national level. Given the primacy of basic science in public funding, the budgets remained nationally organised. At the turn of the century, the member states still account for about 95 percent of expenditures on public civil research and development in the European Union (Banchoff, 2002). This also explains why European networks often gave emerged from non-EU intergovernmental programs. Expenditures in non-EU intergovernmental institutions exceed the current budget of the EU research budget (Banchoff, 2002). Examples of such institutions are the European Centre for Nuclear Research (CERN) and the European Space Agency (ESA).

Hitherto, the EU level of science policy has centred around thematic frameworks that focus primarily on science-based innovation rather than basic science. It is recognised that European science could also benefit from a tighter coordination of national basic science programs, for example, by allowing foreign research groups to compete for national resources. However, recent initiatives to co-ordinate national research programmes and non-EU multilateral programmes through the European Commission have hitherto failed (Banchoff, 2002).

## Internationalisation

The failures of the European Commission to integrate national science policy within a comprehensive logic should not taken to mean that European integration in science is expected to stagnate. In this context, one must carefully distinguish between the integration of science policies (within the political subsystem) and the integration of science itself (within the science subsystem). As part of a more general pattern of ‘internationalisation’ the European research activities may well continue to integrate. An answer to this question can only be provided by empirical research.

The pattern of internationalisation is to some extent exogenous to science policy. The rise in the number of international collaborations in science can also be understood as an organisational consequence of the evolutionary dynamic towards ever greater division-of-labour and specialisation. Other processes often mentioned as contributing to internationalisation are improvements in mobility and ICT and the emergence of English as a world language in science. All these factors have rendered the costs of collaboration much lower than before.

It is, however, important to recognise at this stage of our discussion that there is no theory of scientific collaboration that explains why the number of international collaborations have gone up so drastically over the past twenty years. Such a theory is badly needed, not only for academic purposes, but also in order to systematise research evaluation. Policies can only be evaluated with some degree of precision when factors exogenous to policy can be accounted for, too. As explained in the introduction, we will only deal with empirical indicators in the remainder of the chapter. However, it is important to stress that the development of European indicators is ultimately to be paralleled by the development of theories that explain internationalisation as a historical phenomenon (Wagner, 2002).

## 4. INDICATING INTEGRATION

We will discuss a number of indicators of European integration that have been developed in earlier works (Leydesdorff, 1992; 2000; Frenken, 2002). We proceed in three steps. First, we provide descriptive statistics for the output trends of the European member states. Second, we analyse "systemness'' by testing for the Markov property in the distribution of output data of European member states. In this context, systemness can be understood as a measure of European integration. Third, we analyse collaboration patterns among European member states as expressed by multiple addresses in publication data. European integration can then be tested for by analysing changes in bias of countries to collaborate among each other.

### Descriptive statistics

Output trends provide one with a very basic information on the "performance" of the European system vis-à-vis the rest of the world, and of European member states vis-à-vis each other. *Figure 1* shows performance in terms of percentage of world share of publications, for the European Union in comparison with the U.S.A. and Japan during the period 1980-1998. The European system is indicated for both the European Union of the fifteen current member states (EU) and the European Community of twelve member states.

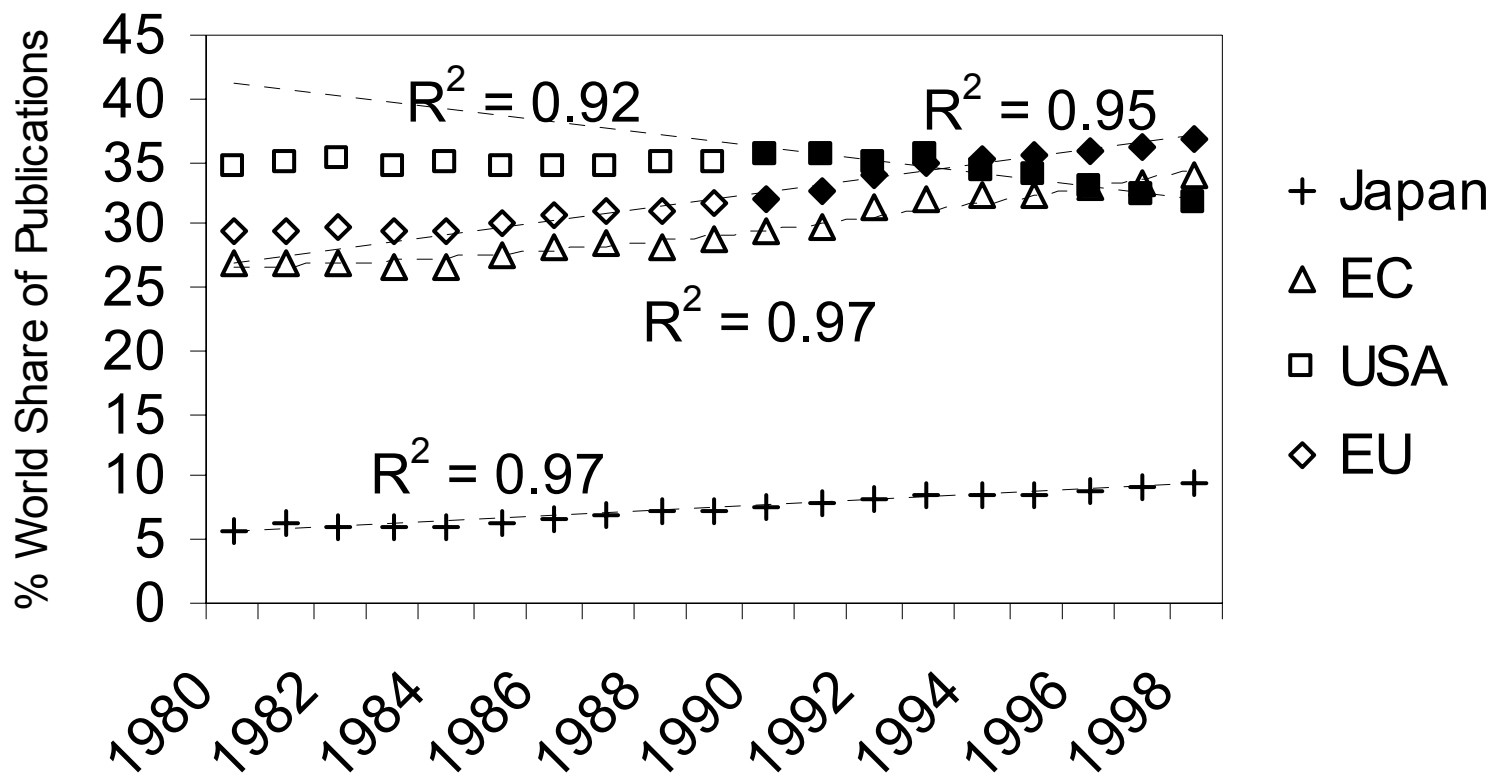

Figure 1
Percentage World Share of Publications for the U.S.A, Japan, and Europe during the 1980s and 1990s.

The figure exhibits, among other things, the relative decline of the U.S. publication system and the advance of the other two major systems. The second polynominal is used for the curve fitting in the case of the EC set in order to highlight how the line potentially deviates from a linear trend ($r > 0.98$): the relative changes of the 1980s seem to be enhanced during the 1990s. Since 1990, the U.S.A. has lost 0.51% per year in terms of its world share of publications ($r > 0.95$), while the EU has gained 0.56% per year ($r > 0.97$).

The overall increase of the share of the European Union during this period (*Figure 1*) was not caused by the R&D systems of the relatively large shares of the UK, France and Germany. As we can see from *Figures 2-4*, European nations differ considerably in their participation in this increase.

The most spectacular growth rates are exhibited by the Italian and Spanish data. As visible from *Figure 2*, this increase is even gaining momentum in the nineties compared to the eighties. During the 1990s, Italy and Spain have grown with 0.14% yearly ($r > 0.95$). Among other countries, depicted in *Figure 3*, some are also increasing their world share of publications, while others like the Netherlands and Sweden have recently witnessed a flattening of output share. The larger countries, shown in *Figure 4*, show no clear trends, though Germany has increased its world share mainly because of the unification in 1991.

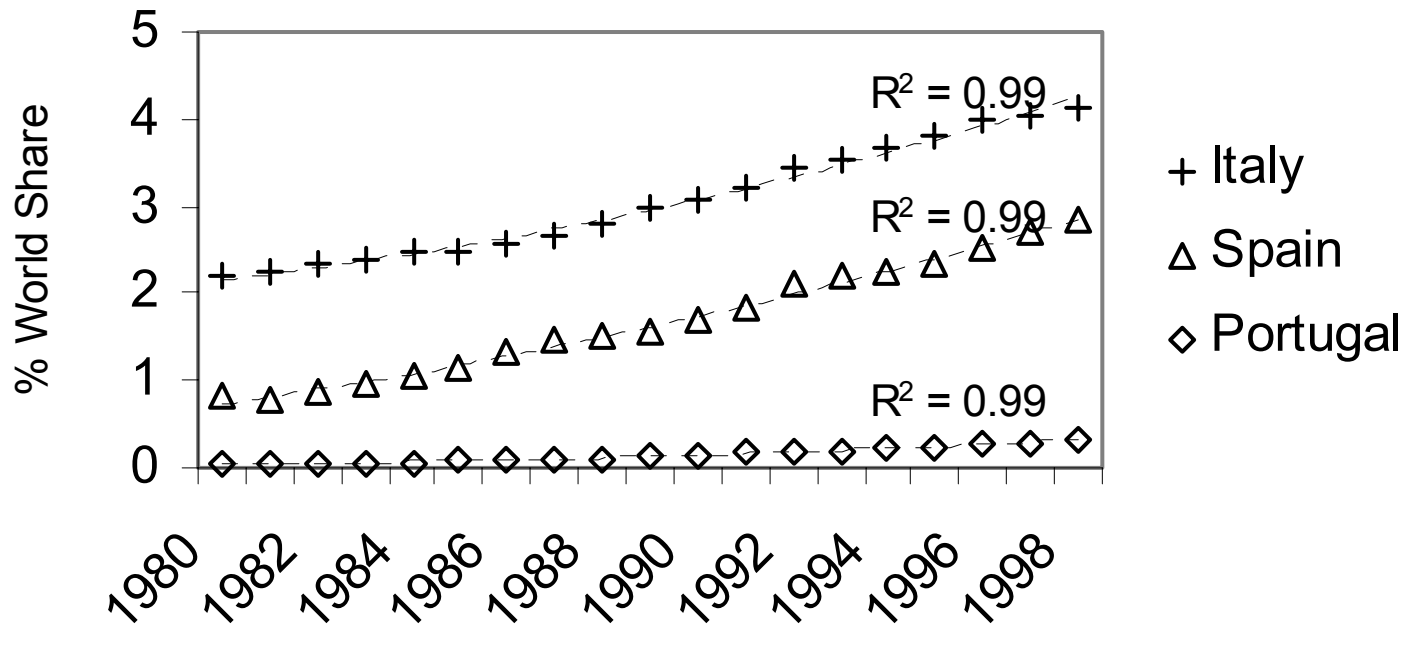

Figure 2
Trend world share of publications for three Southern European countries.

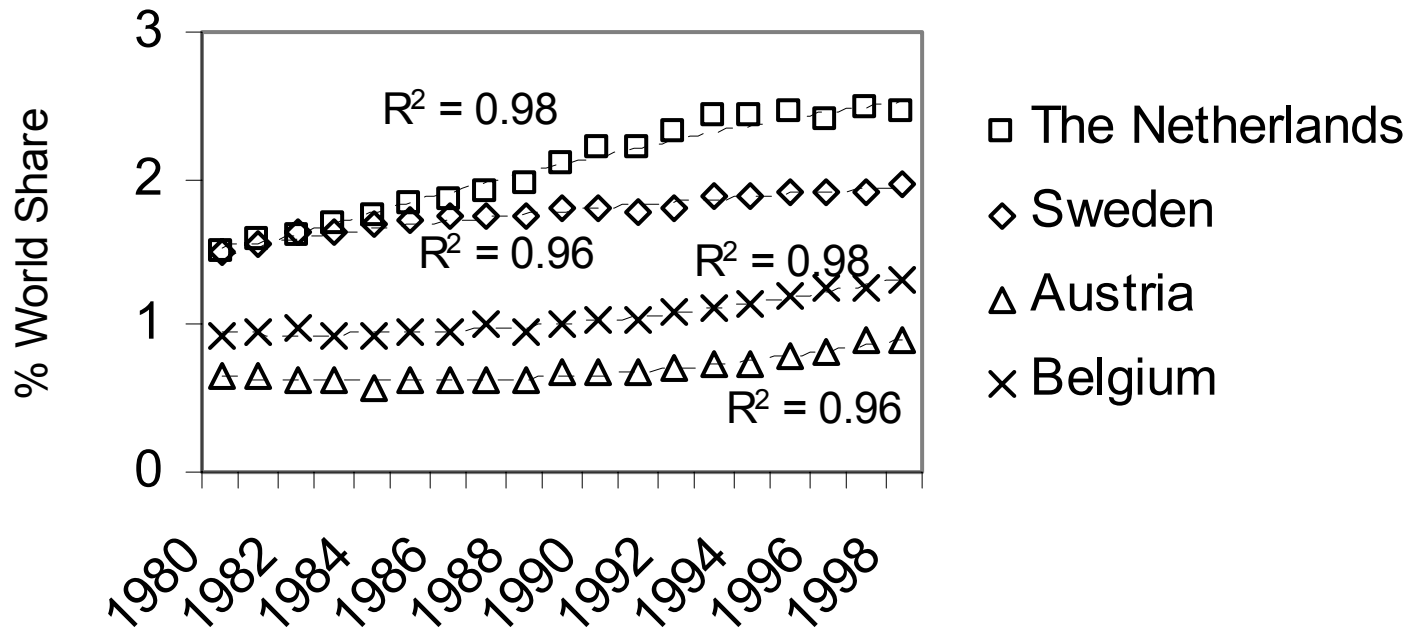

Figure 3
Trend world share of publications for some smaller European countries.

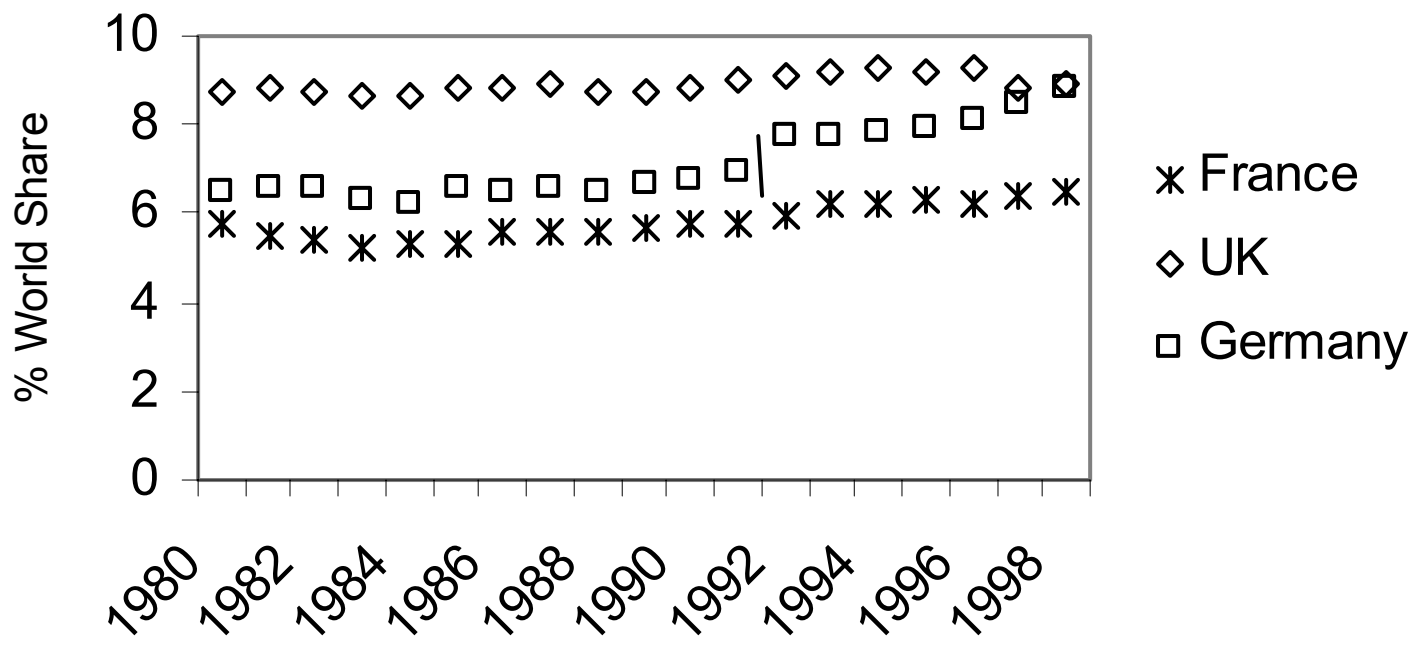

Figure 4
Trend world share of publications for larger European countries.

## Testing for "systemness"

Following work by one of us (Leydesdorff, 1992; 2000), one approach to analysing European integration in scientific research is to view the European system as a *distribution* of output shares of member states. Time-series on scientific output then gives one an evolving yearly distribution of output shares of member states. Using these distributions, information theory can be applied to analyse some aspects of the nature of the underlying evolutionary process.

Let $p_i$ be the share of output of country *i* in year *t* and let $p'_i$ be the share of output of country *i* in year *t+1*. The stability of the distribution can then be measured by looking to what extent the *a priori* distribution at year *t* ($p_1,\ldots, p_{15}$) corresponds to the *a posteriori* distribution at year t+1 ($p'_1,\ldots, p'_{15}$). This is more widely known as the expected information content *I* (Theil, 1967; 1972; Leydesdorff, 1995; Frenken and Leydesdorff, 2000):

$$I(p'/p) = \sum_{i=1}^{15} p'_i \cdot {}^2\log \frac{p'_i}{p_i} \qquad (1)$$

When the shares of all countries remain unchanged during the transition from year *t* to year *t+1*, the I-value would equal zero. This would only be the case when the growth rate of the output share of each country during the transition from year *t* to year *t+1* is exactly the same. As such, the I-value expresses integration in terms of a convergence of growth rates. For any differences in growth rates of countries, the I-value can be shown to be positive (Theil, 1972). The larger the deviations between the resulting distribution of output shares in year *t+1* compared to the previous year *t*, the higher the I-value.

*Figure 5* shows the I-values and the fitting trends for the evolving distribution of the output shares of EU-countries and for the evolving distribution of output shares of the EU plus the US and Japan. The slope in the EU case is negative, while it is slightly positive in the case of the global comparison. This can be considered as an indication of increasing systemness (integration) in the EU data set compared to the global trend since complete integration would mean that $I = 0$.

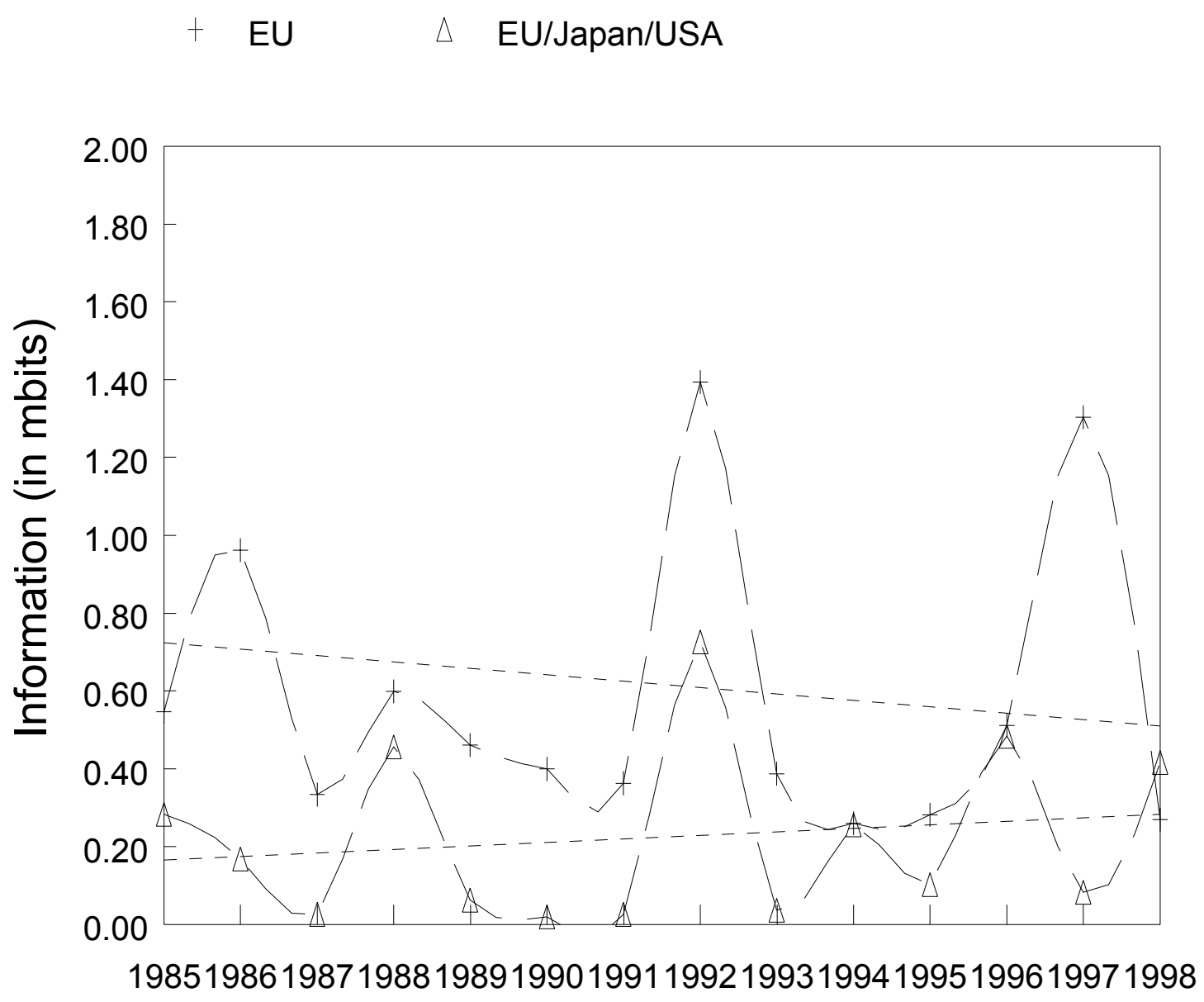

Figure 5
Comparison of the Markov prediction of the EU subset with the set of USA+Japan+EU on a year-to-year basis.

Note that the EU trend is heavily disturbed in 1992. At this time, the effects of the German unification appear in the data as the science system of the former East Germany becomes formally integrated with the data on the former West Germany. Obviously, leaving out the historical event would yield a new fit of the trend line for the EU, which is both smoother and decreasing at a faster rate.

Also note that the I-measure of the expected information content that indicates the stability of an evolving distribution is in itself content-free. It can thus be applied to other social subsystems in the European system can be expressed in a distribution of countries' shares (Leydesdorff and Oomes, 1999).

**Inter-institutional collaboration in research**

The second integration indicator is not based on output shares but on the frequencies of collaborations within and among each European member state. Frenken (2002) defined an inter-institutional collaboration as a pair of different institutional addresses occurring in a publication contained in the Science Citation Index that cover all natural and life sciences (Katz and Martin, 1997).[2] Counting the number of inter-institutional collaborations within and among each country in the European Union, generates a (symmetric) 15x15 matrix containing both intra-national and international collaborations. The number of inter-institutional collaboration between two European member states $i$ ($i=1,..,15$) and $j$ ($j=1,..,15$) as a share of the total number of collaborations is denoted as $q_{ij}$,.

As shown in Frenken (2002), the degree of bilateral integration of country $i$ with respect to country $j$ can then be measured as the difference between the observed share of collaborations $q_{ij}$ and what would be expected from the product of the individual shares $q_i$ and $q_{.j}$. The difference between the observed share and the expected share is measured by the natural logarithm of the division of $q_{ij}$ by the products of $q_{i.}$ and $q_{.j}$:

$$T_{ij} = \ln \frac{q_{ij}}{q_{i.} \cdot q_{.j}} \tag{2}$$

The $T_{ij}$–value is a measure o*f bias*. The value is positive when country $i$ is collaborating with country $j$ more than what is expected from the shares of both countries in all output. The $T_{ij}$–measure is negative when country $i$ is collaborating with country $j$ less than what was expected from their shares. When $i=j$, the measure indicates the bias to collaborate nationally.[3]

---

[2] This definition takes the institutional address as the unit of analysis and not the author. This means that inter-institutional collaboration does not correspond to co-authorship. There are two differences. One person can be associated with more than one institution, which would yield two addresses with only one author in an SCI-record. And, two or more persons can be co-authors associated with the same institution, which would yield one address and two or more authors in an SCI-record. The measurement of scientific collaboration is more thoroughly by Katz and Martin (1997).

[3] An important property of the measure is symmetry of positive and negative bias: that a country collaborating $x$ times more than expected yields value ln($x$) while a country collaborating $x$ times less than expected with another country yields ln(1/$x$). The symmetry of the indicator follows from: ln($x$)=-ln(1/$x$).

Formula (2) gives us a new matrix with all bilateral bias values among each pair of countries. To obtain a single comprehensive integration measure for all fifteen countries, one can use the dependency measure $T$ known as the "mutual information" of a matrix distribution (Theil, 1967; 1972; Langton, 1990; Frenken, 2000):

$$T = \sum_{i=1}^{15} \sum_{j=1}^{15} q_{ij} \cdot \ln \frac{q_{ij}}{q_{i.} \cdot q_{.j}} \qquad (3)^4$$

(Alternatively, one can take the two base logarithm instead of the natural logarithm as to express the indicator in bits as in formula 1, see Theil 1972).

This measure is thus a weighted sum of the bilateral bias-values obtained by formula 2. The larger shares $q_{ij}$ have a correspondingly higher weight in the summation.

It can be shown that the mutual information value $T$ is non-negative for any frequency distribution (Theil, 1967, 1972). When all pairs of countries would collaborate exactly to the extent as expected from the product of their individual shares, all bias values equal zero and the T-value consequently adds up to zero. This would indicate total independency in the matrix distribution, and in our context, perfect integration of the European system. In any other case, the mutual information value will be positive, and the higher the value, the less the countries are integrated in a system. A higher degree of dependency in a matrix distribution thus indicates a lower degree of integration.

What is important to note is that the indicator proposed by Frenken (2002) differs from other integration indicators in that the indicator takes into account both intra-national ($i=j$) and international ($i \neq j$) interactions. In this way, the measure adjusts for size of countries, i.e., for the higher probability of scientists in larger countries to interact with a fellow national citizens compared to scientists in smaller countries. Other measures typically lack this property and thus often indicate that small countries are more internationalised (Frenken, 2002).

The mutual information measure has been applied to the period 1993-2000 and the results are exhibited in *Figure 6*. Clearly, the European Union is integrating as the mutual information falls over time indicating a fall in bias among European member states.

---

[4] For x=0 ; x · ln x = 0.

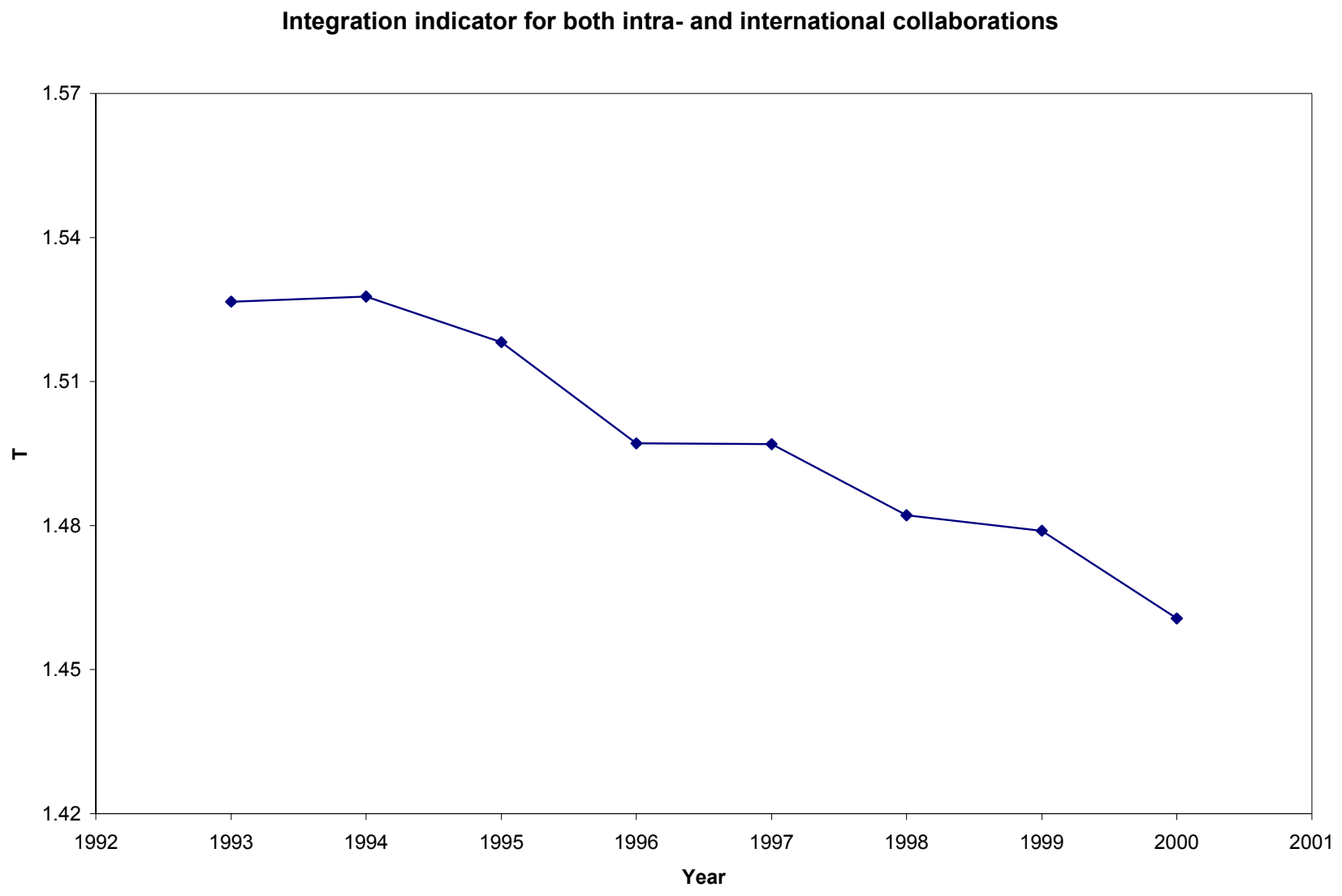

Figure 6
T-values indicating the level of integration of all EU countries

Further analysis has shown that the fall in mutual information indicating European integration is due to a fall in biases among European member states and not a fall in bias to collaborate nationally (Frenken, 2002). This means that the degree of "geographical localisation" does not seem to have decreased. What has changed in the process is that, in so far Europeans collaborate within Europe, they have increasingly lost their bias in choice of partner.

A second observation that has come out of the further analysis is that the largest countries are best integrated. The UK, France and Germany have on average the lowest bias values vis-a-vis other European countries, while smaller countries typically favour collaboration with authors in the larger countries (Frenken, 2002). This result calls for further research into the different 'roles' which small and large countries play within the European science system. The outcome at least suggests that some sort of scale advantages of large countries attract scientists from smaller countries to collaborate with scientists from larger countries. These scale advantages could well be associated with a larger extent of specialisation and a higher budget to invest in expensive research infrastructures.

## 5. DISCUSSION

The integration of the network of coauthorship relations among authors with addresses in European member states does not preclude the conclusion that international coauthorship relations also increase continuously between authors in Europe and authors with addresses outside the EU. However, we could show that the European system exhibits an increasing tendency towards systemness when compared with the relations among Europe, the U.S., and Japan (Figure 5) and that an integration measure applied to the European data shows a steady increase in the internal integration among the EU countries (Figure 6). In addition to this 'Europeanization', 'internationalization' could be shown in Figures 2, 3, and 4 as a development in its own right. A spectacular increase of the visibility of Southern European countries in the international databases during the 1990s (Figure 2) followed upon a similar effect for the smaller countries of Northern Europe during the 1980s (Figure 3). The larger countries (Figure 4) have been mainly stable, with the exception of Germany after its unification in 1991. Germany has become more important as a partner in international collaboration during the 1990s to the extent that at the global level it has taken over functions from the former Soviet-Union (Wagner and Leydesdorff, 2002).

The higher and increasing density of network relations among EU countries can perhaps be compared with trade relations. The network of coauthorship relations is most tightly nit at the national levels, but the European level is an increasingly relevant level. The European programs have been successful by contributing resources to the (nationally integrated) R&D systems which have been in transition towards internationalization to a variable degree.

A first extension of the scientometric studies reported here could be to focus more systematically on the linkages between European science policy, economic competitiveness, and social developments. Note that the beneficial effects of the European integration on scientific knowledge production can be expected *within* the European and national science and higher education systems. It has been shown that international co-authored papers receive significantly more citations than other papers, while it has also been found that international collaboration increases the research productivity of individual scientists (Katz and Martin, 1997). Collaboration not only increases the rate of knowledge production, but also provides a greater diffusion of results and transfer of research skills within the research community. These effects spill over to students through higher education (Katz and Martin, 1997).

These direct effects of European integration in science are important in their own right. Science policy should always first be assessed on meeting the objective to strengthen the scientific knowledge base. However, this conclusion leaves open the question to what extent European networking in science has also contributed to economic and social objectives. These objectives have explicitly been included in European science policies as selection criteria for funding, in particular within the various framework programmes. However, the economic and social impacts remain uncertain as long as European science and technology policies are not supplemented with systematic policy evaluation *ex post*. An important research question within this context would be to investigate scientific disciplines and technological sectors in terms of their sensitivity to European funding and in terms of their effects on science-based innovation and social policies. Also note that, even within disciplines and sectors, variation may arise along geographical lines, rendering some type of policies more effective in particular types of European regions. If science is indeed going through a process of Europeanization, the realization of its potential impacts on economic and social domains may require (supplementary) local policies to account for regional varieties.

A second extension of the scientometric programme could be to analyse networking behaviour among researchers in more detail. In this study we addressed the question of European integration from the perspective of the science system as a whole using the *Science Citation Index* data on all disciplines in natural and life sciences. By doing so, we provided a macro-analysis of European integration and found evidence that this integration process is actually taking place. In science, however, different disciplines are organized in different ways at the meso-level. Science evolves mainly through self-organising processes of communication and collaboration within 'invisible colleges' that form disciplines and specialties at national and international levels. Therefore, explanatory research should go beyond indicator analysis and replace the institutional unit of analysis (e.g., national addresses) with an intellectual unit of analysis (e.g., journal sets). Can a European level of self-organization be made visible in the case of techno-sciences like 'biotechnology' (Leydesdorff and Heimeriks, 2001)? Can Information and Communication Technology be considered a relevant (European) unit of analysis? Theorising about the determinants and effects of collaboration should begin at delineating scientific disciplines and to take into account their specificities. A research program with a focus on rationales and dynamics of collaboration in different disciplines is currently underway (Wagner, 2002).

**Dr. Koen Frenken** (PhD, Amsterdam and Grenoble) is lecturer at the Faculty of Geographical Sciences and researcher at the Urban Research centre Utrecht (URU) at the Utrecht University. He has published on evolutionary economics, technological innovation, and the geography of knowledge production. E-mail: k.frenken@geog.uu.nl. Home-page : http://econ.geog.uu.nl/frenken/frenken.html

**Dr. Loet Leydesdorff** (PhD, Amsterdam) is senior lecturer at the Faculty of Social and Behavioural Sciences and a senior researcher at the Amsterdam School of Communications Research (ASCoR) at the University of Amsterdam. He has published extensively on sociology, scientometrics, philosophy of science, and systems theory. E-mail: loet@leydesdorff.net. Home-page : http://www.leydesdorff.net

**Bibliography**

Banchoff, T. (2002). 'Institutions, inertia and European Union research policy.' *Journal of Common Market Studies* **40**(1): 1-21.

Bush, V. (1945). *The Endless Frontier: A Report to the President*. Reprinted New York, Arno Press, 1980.

Collins, H.M. (1985). 'The possibilities of science policy.' *Social Studies of Science* **15**: 554-558.

Frenken, K. (2000). 'A complexity approach to innovation networks. The case of the aircraft industry (1909-1997).' *Research Policy* **29**(2): 257-272.

Frenken, K. (2002). 'A new indicator of European integration and an application to collaboration in scientific research.' *Economic Systems Research* **14**(4): forthcoming.

Frenken, K., L. Leydesdorff (2000). 'Scaling trajectories in civil aircraft (1913-1997).' *Research Policy* **29**(3): 331-348.

Katz, S., B.R. Martin (1997). 'What is research collaboration?' *Research Policy* **26**(1): 1-18.

Langton, C.G. (1990). 'Computation at the edge of chaos. Phase-transitions and emergent computation.' *Physica D* **42**(1-3): 12-37.

Leydesdorff, L. (1992). 'The impact of EC science policies on the transnational publication system.' *Technology Analysis & Strategic Management* **4**: 279-298.

Leydesdorff, L. (1995). *The Challenge of Scientometrics. The Development, Measurement, and Self-Organization of Scientific Communications*. Leiden, DSWO Press, Leiden University.

Leydesdorff, L. (2000). 'Is the European Union becoming a single publication system?' *Scientometrics* **47**(2): 265-280.

Leydesdorff, L. (2003). 'Scientometrics indicators and the evaluation of research.' *La Revue pour l'Histoire de la Recherche*: forthcoming.

Leydesdorff, L., S.E. Cozzens (1993). 'The delineation of specialities in terms of journals using the dynamic journal set of the SCI.' *Scientometrics* **26**(1): 135-156.

Leydesdorff, L., G. Heimeriks (2001). 'The self-organization of the European information society: The case of "biotechnology".' *Journal of the American Society for Information Science & Technology JASIST* **52**(14): 1262-1274.

Leydesdorff, L., N. Oomes (1999). 'Is the European monetary system converging to integration?' *Social Science Information* **38**(1): 57-86.

Luukkonen, T. (1998). 'The difficulties in assessing the impact of EU framework programmes.' *Research Policy* **27**(6): 599-610.

Moed, H.F., W.J.M. Burger, J.G. Frankfort, A.F.J. Van Raan (1985). 'The use of bibliometric data for the measurement of university research performance.' *Research Policy* **14**(131-149).

Narin, F., N. Elliott (1985). 'Is technology becoming science?' *Scientometrics* **7**(369-381).

Price, D.d.S. (1963). *Little Science, Big Science*. New York, Columbia University Press.

Theil, H. (1967). *Economics and Information Theory*. Amsterdam, North-Holland.

Theil, H. (1972). *Statistical Decomposition Analysis*. Amsterdam, North-Holland.

Van den Besselaar, P., L. Leydesdorff (1996). 'Mapping change in scientific specialties: A scientometric reconstruction of the development of artificial intelligence.' *Journal of the American Society for Information Science* **47**(6): 415-436.

Wagner, C.S. (2002). *International Linkages: Is Collaboration Creating a New Dynamic for Knowledge Creation in Science?*, Manuscript, Amsterdam School of Communications Research, University of Amsterdam.

Wagner, C.S., L. Leydesdorff (2002). *Mapping the Global Network Using International Co-Authorships: A Comparison of 1990 and 2000*, Manuscript, Amsterdam School of Communications Research, University of Amsterdam.

Weingart, P. (1991). *Die Wissenschaft in osteuropäischen Ländern im internationalen Vergleich-eine quantitative Analyse auf der Grundlage wissenschaftsmetrischer Indikatoren*. Bielefeld, Kleine Verlag.

Whitley, R.R. (1984). *The Intellectual and Social Organization of the Sciences*. Oxford, Oxford University Press.